\documentclass[preprint,showpacs,preprintnumbers,amsmath,amssymb,floatfix,endfloats*]{revtex4}

\usepackage{graphicx}
\usepackage{dcolumn}
\usepackage{bm}
\usepackage{amsfonts}

\DeclareMathAlphabet{\mathsfsl}{OT1}{cmr}{bx}{it}
\begin{document}
\title{Heterogeneous relaxation dynamics in amorphous materials under cyclic loading}
\author{Nikolai V. Priezjev}
\affiliation{Department of Mechanical Engineering, Michigan State
University, East Lansing, Michigan 48824}
\date{\today}
%
\begin{abstract}

Molecular dynamics simulations are performed to investigate
heterogeneous dynamics in amorphous glassy materials under
oscillatory shear strain.     We consider three-dimensional binary
Lennard-Jones mixture well below the glass transition temperature.
The structural relaxation and dynamical heterogeneity are quantified
by means of the self-overlap order parameter and the dynamic
susceptibility.  We found that at sufficiently small strain
amplitudes, the mean square displacement exhibits a broad
sub-diffusive plateau and the system undergoes nearly reversible
deformation over about $10^4$ cycles.  Upon increasing strain
amplitude, the transition to the diffusive regime occurs at shorter
time intervals and the relaxation process involves intermittent
bursts of large particle displacements.  The detailed analysis of
particle hopping dynamics and the dynamic susceptibility indicates
that mobile particles aggregate into clusters whose sizes increase
at larger strain amplitudes.   Finally, the correlation between
particle mobilities in consecutive time intervals demonstrates that
dynamic facilitation becomes increasingly pronounced at larger
strain amplitudes.

\end{abstract}



\maketitle

\section{Introduction}

Understanding the relationship between atomic structure and
mechanical properties in amorphous materials is important in many
current applications and emerging technologies~\cite{Ediger12}. In
contrast to crystalline solids, where the plastic deformation is
governed by dislocations, it was originally found that the plastic
activity in amorphous materials is controlled by the localized shear
transformations~\cite{Argon79}, which were more recently studied
using computer simulations~\cite{Falk98,Maloney06,Tanguy06,Delogu08}
and directly visualized in experiments on colloidal
glasses~\cite{Weitz07,Weeks10} and foams~\cite{Dennin08}. Under the
applied strain, the deformation of amorphous materials is determined
by the cooperative organization of irreversible rearrangements of
small clusters of particles, which could be triggered by the
nonlocal redistribution of elastic
stress~\cite{Maloney04,Demkowicz05,Lemaitre09,Martens11}.  The
sequence of such plastic events can lead to an avalanche process
characterized by a power-law scaling of the average stress or energy
drops with the system size~\cite{Jacobsen07,Procaccia09}.  Another
notable examples of systems with intermittent, spatiotemporal
heterogeneous dynamics include the Barkhausen crackling noise in
magnets subject to an external magnetic field~\cite{Colaiori08} and
the driven block-spring model in the theory of self-organized
criticality~\cite{Bak87,BakNature}.

In recent years, computer simulations have become an increasingly
important tool for studying slow particle dynamics in molecular
liquids near the glass transition at thermal
equilibrium~\cite{Berthier11}.       A common observation is that
the mean square displacement of individual particles is reduced upon
approaching the glass transition temperature from above, which also
results in broadening of the sub-diffusive plateau that separates
the ballistic and diffusive regimes.     More importantly, however,
is that the particle mobility in the sub-diffusive regime can be
significantly different from the average value; and, in addition,
the particles with similar mobility become spatially correlated,
thus leading to dynamic
heterogeneity~\cite{Cavagna09,Donati98,Harrowell99}. The spatial
fluctuations of mobile regions are characterized by the four-point
dynamic correlation function, and they can be efficiently measured
by computing the variance of the self-overlap order parameter, or
the dynamic susceptibility~\cite{Glotzer00,Glotzer03}. While no
obvious changes in spatial correlations of particle positions are
detected near the glass transition, the dynamic susceptibility
provides an estimate of the number of particles involved in the
correlated motion.     It is now well recognized that when a liquid
is cooled toward the glass transition temperature, the peak value of
the dynamic susceptibility increases, indicating that dynamics
becomes spatially increasingly correlated~\cite{Berthier11}.   More
recently, it was shown that in the presence of steady shear flow,
the dynamics of supercooled liquids is more homogeneous as the shear
flow reduces the dynamic correlation length and the lifetime of
dynamical heterogeneity~\cite{Yamamoto12}.

The microscopic mechanism of structural relaxation in glassy
materials is governed by spatially extended domains of fast moving
particles that can be identified from the analysis of individual
particle trajectories.     Hopping particle dynamics was recently
studied in a number of systems, i.e., binary mixtures below the
glass transition~\cite{Vollmayr04}, supercooled liquids at thermal
equilibrium~\cite{BiroliPRL10}, dense granular
media~\cite{BiroliPRL09, BiroliEPL10}, and actively deformed polymer
glasses~\cite{Rottler10}.     In each case, a particle trajectory
was decomposed into a series of segments, where motion takes place
inside a cage, separated by fast cage jumps. Of particular
importance is the analysis of cage jumps and their spatio-temporal
clusterization performed in the cyclic shear~\cite{BiroliPRL09} and
the fluidized bed~\cite{BiroliEPL10,GlotzerNat07} experiments of
two-dimensional granular media.    In both experiments, it was found
that the major structural relaxation events are well correlated with
the bursts of cage jumps.    Furthermore, these cage jumps tend to
aggregate into clusters, whose sizes are approximately power-law
distributed~\cite{BiroliPRL09,BiroliEPL10}.      In turn, several
clusters might dynamically facilitate each other and form
avalanches, which propagate along the soft regions of the
system~\cite{ BiroliPRL09,BiroliEPL10,BiroliPRL10}.       One of the
motivations of the present study is to examine the collective motion
of particles and dynamic facilitation in periodically deformed
three-dimensional amorphous materials.

In this paper, molecular dynamics simulations are employed to study
the relaxation dynamics in the Kob-Andersen Lennard-Jones binary
mixture model at a finite temperature well below the glass
transition.    The three-dimensional system is periodically strained
over many cycles, probing regions with low instability thresholds,
which leads to intermittent localized rearrangements of particles.
We find that at sufficiently large strain amplitudes the particle
dynamics is purely diffusive, while at lower amplitudes, the mean
square displacement develops an extended sub-diffusive plateau. The
analysis of the dynamical susceptibility and particle hopping
dynamics reveals the spatial heterogeneity of structural relaxation.

The rest of the paper is organized as follows.   The details of
molecular dynamics simulations are described in the next section.
The results for the particle diffusion, hopping dynamics, and
microstructure of clusters of mobile particles, as well as the
analysis of the two and four-point correlation functions are
presented in Sec.\,\ref{sec:Results}.   The conclusions are given in
the last section.

\section{Molecular dynamics simulation model}
\label{sec:MD_Model}

\begin{figure}[t]
\includegraphics[width=10.cm,angle=0]{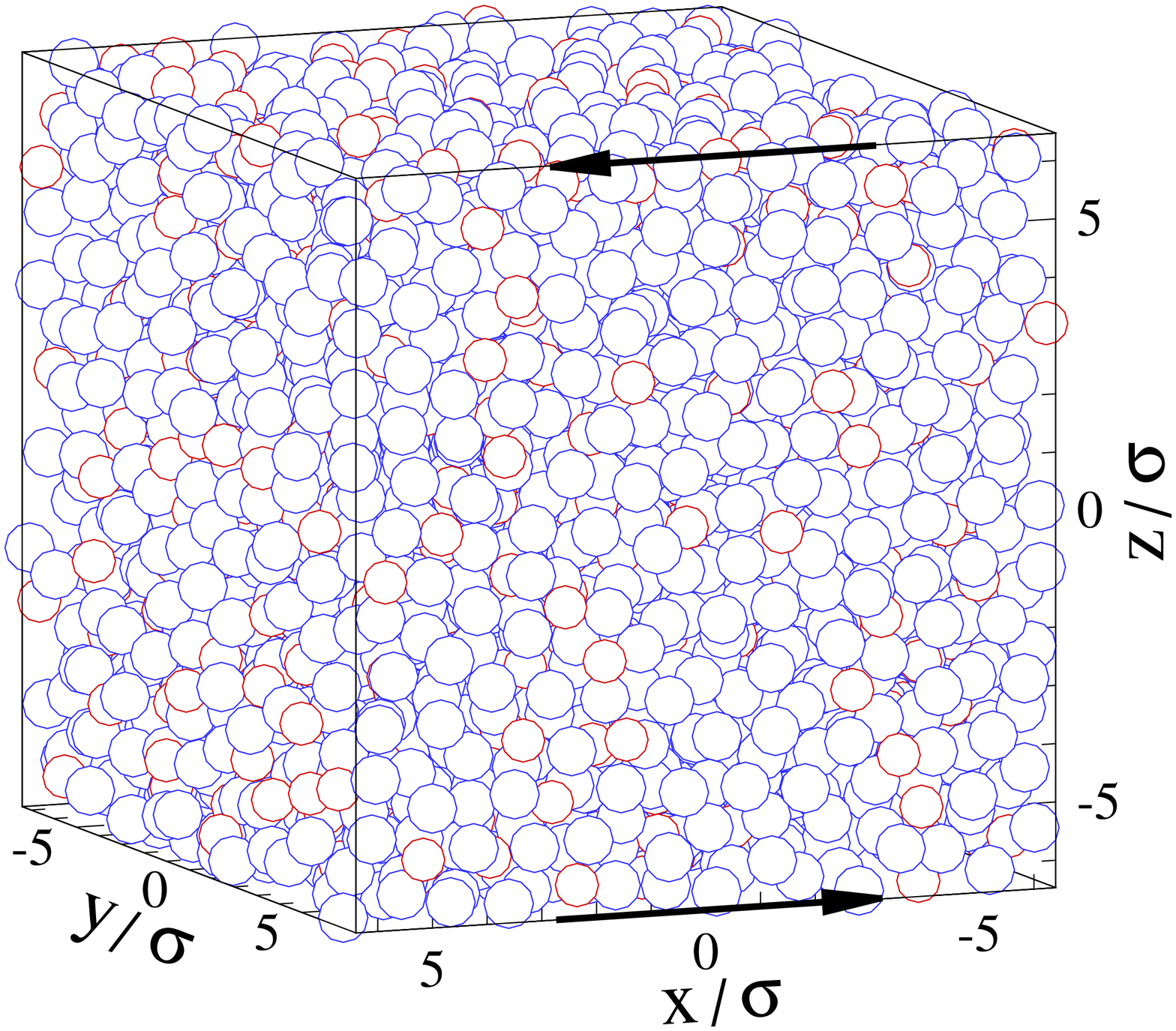}
\caption{(Color online) A snapshot of instantaneous positions of
particles $A$ (blue circles) and $B$ (red circles) at mechanical
equilibrium.   The particle sizes are not drawn to scale.   During
the oscillatory motion, the periodic shear strain was applied in the
$xz$ plane (indicated by the arrows). } \label{fig:snapshot_system}
\end{figure}

The three-dimensional amorphous material is modeled as the
Kob-Andersen binary (80:20) Lennard-Jones mixture with non-additive
interaction parameters that prevent crystallization~\cite{KobAnd95}.
The snapshot of the equilibrated system which consists of $N_p=2940$
particles is presented in Fig.\,\ref{fig:snapshot_system}. In this
model, any two particles $\alpha,\beta=A,B$ interact via the
pairwise Lennard-Jones (LJ) potential
\begin{equation}
V_{\alpha\beta}(r)=4\,\varepsilon_{\alpha\beta}\,\Big[\Big(\frac{\sigma_{\alpha\beta}}{r}\Big)^{12}\!-
\Big(\frac{\sigma_{\alpha\beta}}{r}\Big)^{6}\,\Big],
\label{Eq:LJ_KA}
\end{equation}
where    $\varepsilon_{AA}=1.0$,     $\varepsilon_{AB}=1.5$,
$\varepsilon_{BB}=0.5$,    $\sigma_{AB}=0.8$,   $\sigma_{BB}=0.88$,
and $m_{A}=m_{B}$.       The cutoff radius is taken to be twice the
minimum position of the LJ potential
$r_{c,\,\alpha\beta}=2.245\,\sigma_{\alpha\beta}$~\cite{Varnik04,Varnik06}.
In what follows, the units of length, mass, and energy are set to be
$\sigma=\sigma_{AA}$, $m=m_{A}$, and $\varepsilon=\varepsilon_{AA}$,
and, correspondingly, the unit of time is defined as
$\tau=\sigma\sqrt{m/\varepsilon}$.      The equations of motion were
solved numerically using the fifth-order Gear predictor-corrector
algorithm~\cite{Allen87} with a time step $\triangle
t_{MD}=0.005\,\tau$.

All simulations were performed at a constant volume with the total
density $\rho=\rho_{A}+\rho_{B}=1.2\,\sigma^{-3}$ and temperature
$0.1\,\varepsilon/k_B$, where $k_B$ is the Boltzmann constant. This
temperature is well below the value $0.45\,\varepsilon/k_B$ at which
the computer glass transition is detected~\cite{KobAnd95}. The
constant temperature was maintained by rescaling the velocity
component in the $\hat{y}$ direction (perpendicular to the plane of
shear).   As indicated in Fig.\,\ref{fig:snapshot_system}, the
system dimensions are measured $L_x\,{=}\,12.81\,\sigma$,
$L_y\,{=}\,14.79\,\sigma$, and $L_z\,{=}\,12.94\,\sigma$.   In order
to simulate homogeneous, time-dependent shear strain, the
Lees-Edwards periodic boundary conditions~\cite{Allen87} were
implemented with the SLLOD equations of motion~\cite{Evans92}.  It
should be mentioned that in contrast to the boundary-driven shear
algorithms, the spatially homogeneous shear strain prevents the
formation of shear bands~\cite{VarnikBBB03}.

The time-periodic shear strain was imposed (in the $xz$ plane) by
varying the strain rate as a function of time
$\dot{\gamma}(t)=\dot{\gamma}_{0}\,\textrm{cos}(\omega t)$, where
$\omega$ is the oscillation frequency and $\dot{\gamma}_{0}$ is the
strain rate amplitude.    We define the strain amplitude as a ratio
of the strain rate amplitude over the frequency, i.e.,
$\gamma_{0}=\dot{\gamma}_{0}/\omega$.   For the results reported in
this paper, the oscillation frequency was set $\omega\tau=0.02$,
with the corresponding period $T=2\pi/\omega=314.16\,\tau$; and the
strain amplitude was varied in the range $\gamma_{0}\leqslant 0.08$.

The system was first equilibrated for about $5\times10^6$ MD steps
at a constant volume and temperature $1.2\,\varepsilon/k_B$ in the
absence of shear, and then gradually quenched to the final
temperature $0.1\,\varepsilon/k_B$ with steps of
$0.1\,\varepsilon/k_B$.   After the oscillatory shear strain was
applied, the first $2\times10^7$ MD steps were discarded to avoid
quench-rate and aging effects.   During the oscillatory motion, the
measurements of particle positions were taken every back and forth
cycle when strain is zero.    The data were accumulated over
$12,000$ cycles (about $7.5\times10^8$ MD steps) at each strain
amplitude, and the post-processing analysis of particle trajectories
was performed in six independent systems.

\section{Results}
\label{sec:Results}

At the studied temperature and density, an equilibrated model glass
in the absence of deformation is characterized by the amorphous
liquid-like molecular structure where most of the atoms remain in
cages formed by their neighbors on the time scale accessible to
computer simulations~\cite{KobAnd95}.   A typical steady shear
stress--strain response involves an elastic deformation at strains
below a few percent and a yield stress that depends on the physical
aging and strain rate~\cite{Varnik04,Rottler05}.   During the
elastic and plastic deformations, the atoms can undergo non-affine,
irreversible displacements, which, depending on the strain, might
form cascades spanning a considerable fraction of the
system~\cite{BarratCh8}.   Thus, instead of temperature, the
consecutive irreversible displacements of atoms are governed by the
strain rate as a control parameter~\cite{Tsamados10}.    In the
present study, the amorphous material is periodically deformed, and
the particle positions are saved every cycle at zero strain.   In
such a setup, the affine deformation field, which is present at
steady shear strain, is zero; and, therefore, any irreversible
particle rearrangements will contribute to the structural relaxation
of the material.

\begin{figure}[t]
\includegraphics[width=12.cm,angle=0]{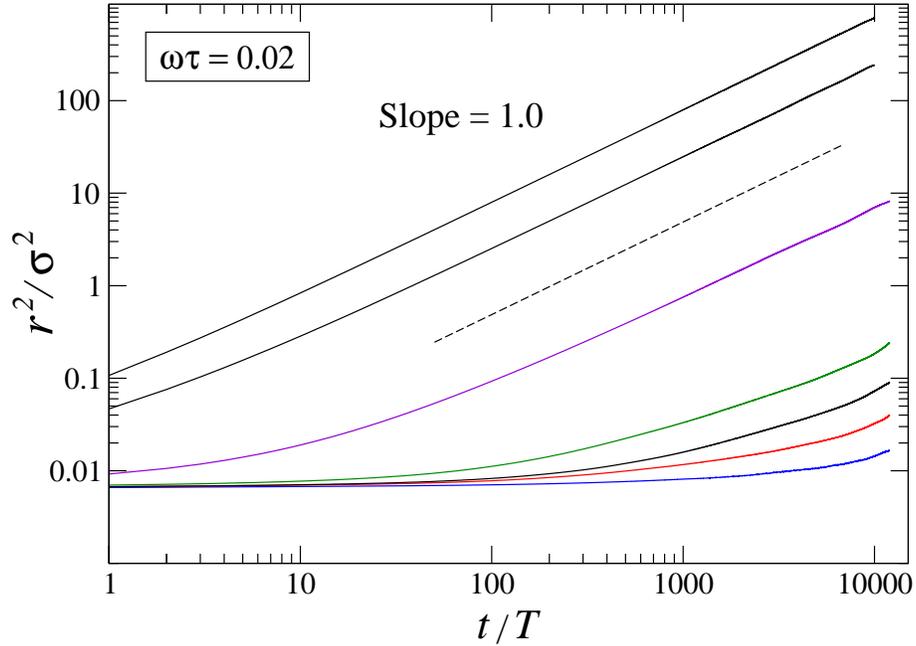}
\caption{(Color online) The mean square displacement of $A$ and $B$
particles as a function of time for the oscillation frequency
$\omega\tau=0.02$ and period $T=2\pi/\omega=314.16\,\tau$.   The
strain amplitudes from bottom to top are
$\gamma_{0}=0.02,~0.03,~0.04,~0.05,~0.06,~0.07,~\text{and}~0.08$.
The dashed line with unit slope is plotted for reference. }
\label{fig:msd_time_omega}
\end{figure}

The mean square displacement (MSD) averaged over both $A$ and $B$
particles is plotted in Fig.\,\ref{fig:msd_time_omega} as a function
of time for the frequency $\omega\tau=0.02$ and various strain
amplitudes.       To compute the displacement, the position of the
system center of mass was subtracted from the position of each
particle.      It can be observed in Fig.\,\ref{fig:msd_time_omega}
that at small strain amplitudes, $\gamma_{0} \leqslant 0.06$, the
MSD curves exhibit a broad sub-diffusive plateau, which becomes more
pronounced at smaller strain amplitudes, and a gradual crossover to
diffusive motion.       In contrast, at larger strain amplitudes,
$\gamma_{0} \geqslant 0.07$, the sub-diffusive regime is absent, and
the slope of the MSD curves becomes equal to one at times $t\gtrsim
10\,T$ as indicated by the straight dashed line in
Fig.\,\ref{fig:msd_time_omega}.     At the largest time interval
$t=1.2\times10^4\,T$, the particle displacement is still about the
cage size for the strain amplitude $\gamma_{0}=0.02$, which implies
that during the periodic deformation the system dynamics is nearly
reversible.    Note also that the ballistic regime is not observed
in any of the MSD curves as it occurs at times much smaller than the
oscillation period $T=314.16\,\tau$.


The appearance of the extended sub-diffusive plateau in the MSD
curves reported in Fig.\,\ref{fig:msd_time_omega} suggests that the
particle dynamics might be spatially heterogeneous on the length
scales of about the cage size.   The structural relaxation in
amorphous materials is commonly quantified via the self-correlation
function, which is defined as follows:
\begin{equation}
Q_s(a,t)=\frac{1}{N_{p}} \sum_{i=1}^{N_{p}}\text{exp}\Big( -
\frac{\Delta\mathbf{r}_{i}(t)^2}{2\,a^{2}} \Big),
\label{Eq:self_corr}
\end{equation}
where
$\Delta\mathbf{r}_{i}(t)=\mathbf{r}_{i}(t_0+t)-\mathbf{r}_{i}(t_0)$
is the displacement vector of the $i$th particle, $t$ is the lag
time, and $a$ is the probed length scale~\cite{BiroliPRL05}.   In
turn, the extent of dynamical heterogeneity is measured by the
four-point correlation function, or the dynamical susceptibility,
which is computed as the variance of the self-correlation function:
\begin{equation}
\chi_4 (a,t) = N_{p}\,\big[ \langle Q_s(a,t)^2 \rangle - \langle
Q_s(a,t) \rangle^2 \big],    \label{Eq:four_point_corr}
\end{equation}
where the brackets $\langle \cdot \rangle$ denote averaging over all
particles and initial times~\cite{Glotzer00}.    At a given time
lag, the correlation function $\chi_4(a,t)$ provides an estimate of
the number of particles involved in a cooperative displacement over
the length scale $a$~\cite{Berthier11}.    At some intermediate time
and length scales, the function $\chi_4(a,t)$ usually displays a
maximum indicating the largest spatial correlation between localized
particles~\cite{Glotzer00}.

\begin{figure}[t]
\includegraphics[width=12.cm,angle=0]{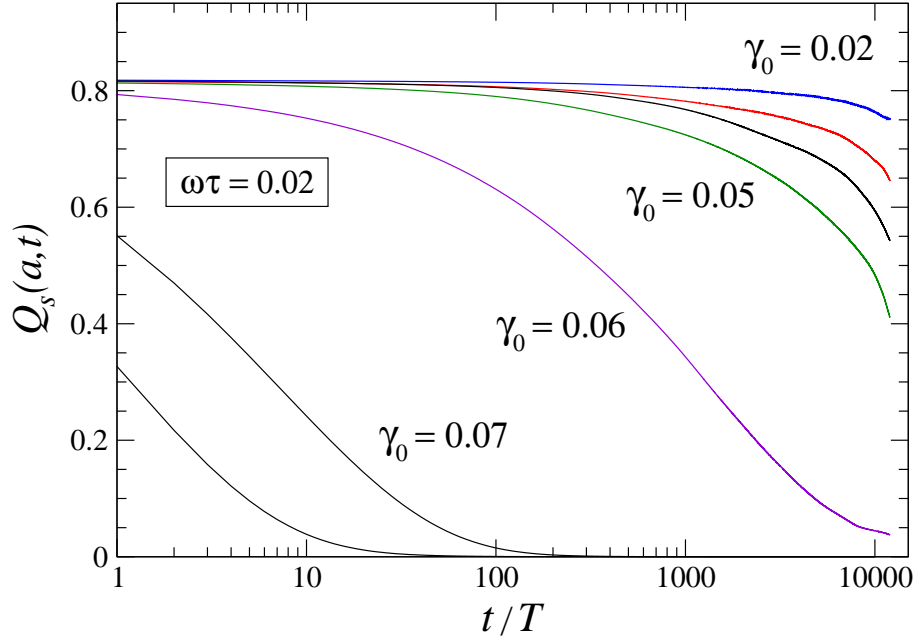}
\caption{(Color online)  The self-correlation function $Q_s(a,t)$
defined by Eq.\,(\ref{Eq:self_corr}) for the oscillation frequency
$\omega\tau=0.02$ and period $T=2\pi/\omega=314.16\,\tau$.    The
probed length scale is $a=0.12\,\sigma$.    The strain amplitudes
from top to bottom are
$\gamma_{0}=0.02,~0.03,~0.04,~0.05,~0.06,~0.07,~\text{and}~0.08$. }
\label{fig:q28_time_omega_0.02}
\end{figure}
\begin{figure}[t]
\includegraphics[width=12.cm,angle=0]{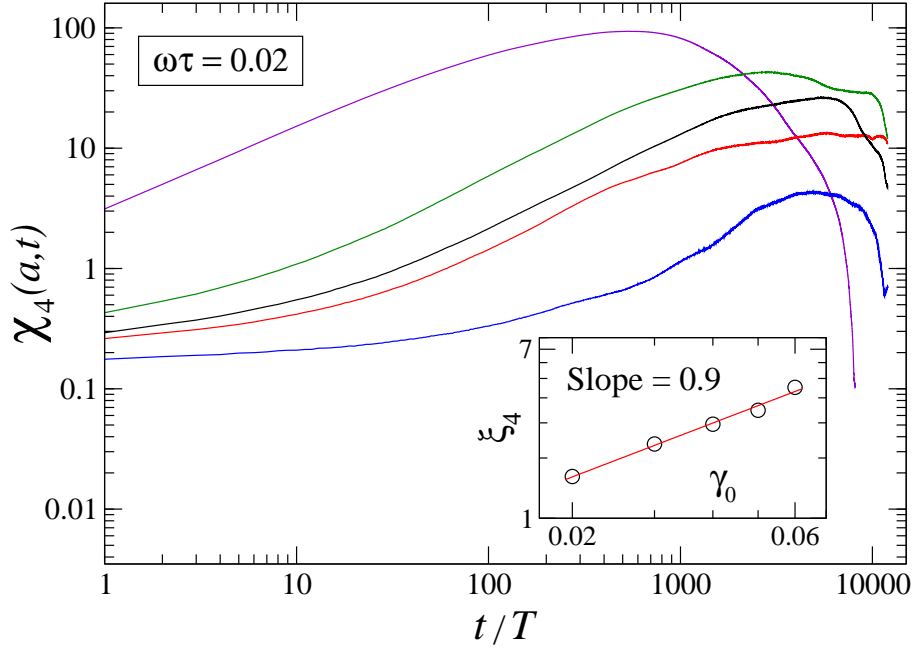}
\caption{(Color online)  The dynamic susceptibility $\chi_4(a,t)$
for the oscillation frequency $\omega\tau=0.02$ and
$a=0.12\,\sigma$.     The strain amplitudes from bottom to top are
$\gamma_{0}=0.02,~0.03,~0.04,~0.05,~0.06$.  The oscillation period
is $T=314.16\,\tau$.    The inset shows the dynamic correlation
length $\xi_{4}=[\chi_4^{\text{max}}(t)]^{1/3}$ as a function of the
strain amplitude $\gamma_{0}$.     The red line with a slope $0.9$
is the best fit to the data.   } \label{fig:ksi28_time_omega_0.02}
\end{figure}

The time dependence of the self-correlation function $Q_s(a,t)$ is
illustrated in Fig.\,\ref{fig:q28_time_omega_0.02} when
$\omega\tau=0.02$ and the parameter $a$ is slightly larger than the
cage size, i.e., $a=0.12\,\sigma$.       As is evident, the
correlation function $Q_s(a,t)$ decays faster at larger strain
amplitudes.      Note that at smaller strain amplitudes $\gamma_{0}
\leqslant 0.05$, the system is not fully relaxed even at the largest
time interval $t=1.2\times10^4\,T$.    On the other hand, at the
smallest time interval $t=T$, the function $Q_s(a,t)$ is less than
$1.0$ because of the thermal vibrations inside a cage.  Further, the
dynamic susceptibility $\chi_4(a,t)$ is shown in
Fig.\,\ref{fig:ksi28_time_omega_0.02} for the same parameters
$\omega\tau=0.02$ and $a=0.12\,\sigma$.     It can be observed that
the correlation function $\chi_4(a,t)$ exhibits a pronounced peak,
whose magnitude increases at larger strain amplitudes, indicating
progressively larger size of dynamically correlated regions.    Upon
increasing strain amplitude, the peak is displaced to smaller times,
which is consistent with the onset of the diffusive regime in MSD
curves reported in Fig.\,\ref{fig:msd_time_omega}.     Finally,
assuming that the correlated regions are compact, the dynamic
correlation length $\xi_{4}$ can be estimated from the peak value of
$\chi_4(a,t)$ at $a=0.12\,\sigma$.        The inset in
Fig.\,\ref{fig:ksi28_time_omega_0.02} shows $\xi_{4}$ as a function
of the stain amplitude.    It is seen that the data are well fitted
by the power-law function with the exponent of about $0.9$ (straight
red line in Fig.\,\ref{fig:ksi28_time_omega_0.02}).    One should
keep in mind, however, that the dynamic susceptibility is an
averaged quantity, which measures mean square fluctuations of the
number of mobile particles, and, thus, it does not completely
describe the microscopic mechanism of structural relaxation.     We
next perform a more detailed analysis of the particle hopping
dynamics and the local microstructure of clusters of mobile
particles.


At sufficiently low temperature and small strain, a typical particle
trajectory in a glassy material consists of rapid hopping events
separated by the rattling motion within a cage.   Hence, the hopping
dynamics is controlled by the cage-to-cage jumps, which, in
practice, can be identified by a numerical algorithm recently
introduced by Candelier\,\textit{et al.}~\cite{BiroliPRL09}.    In
essence, this algorithm is based on the spatial separation of two
consecutive segments of a particle trajectory. More specifically,
the measure of the distance separating two segments is defined by
the product of the root mean square distances between all points
within the segments to the center of mass of the other
segment~\cite{BiroliPRL09}.   Furthermore, the effective distance
between two segments is normalized by a factor that counterbalances
large fluctuations arising from averaging over short segments.
During the iterative procedure, the cage jumps are detected if the
effective distance is greater than the typical cage size; and the
whole trajectory is consecutively divided into a number of segments
where the particle motion takes place inside a cage.   This
algorithm was successively applied to identify cage jumps in
two-dimensional granular systems under cyclic
loading~\cite{BiroliPRL09}, in the fluidized bed
experiment~\cite{BiroliEPL10}, and in supercooled liquids at
mechanical equilibrium~\cite{BiroliPRL10}.


In the present study, the cage detection algorithm was used to
analyze three-dimensional trajectories of individual particles as
follows.    First, we choose a subset of points in the particle
trajectory, divide the subset in two adjacent segments, and then
compute the effective distance separating these two segments.   If
the effective distance between any two adjacent segments within the
subset is less than the cage size, $r_c=0.1\,\sigma$, then the
particle was considered being in the cage during the time interval
defined by the subset.    For every particle, the procedure was
repeated for all time intervals greater than $10\,T$ and less than
$100\,T$ and all initial times.   As a result, all particle
trajectories were decomposed into successive cages separated by cage
jumps, which typically consist of several consecutive points each. A
visual examination of the trajectories revealed two types of jumps;
namely, reversible, where a particle jumps back and forth between
the averaged positions, and irreversible, where a particle
permanently escapes its cage.   This is consistent with the results
of previous MD studies of glassy systems at
equilibrium~\cite{Hansen88,Wahnstrom91,Vollmayr04}.

\begin{figure}[t]
\includegraphics[width=12.cm,angle=0]{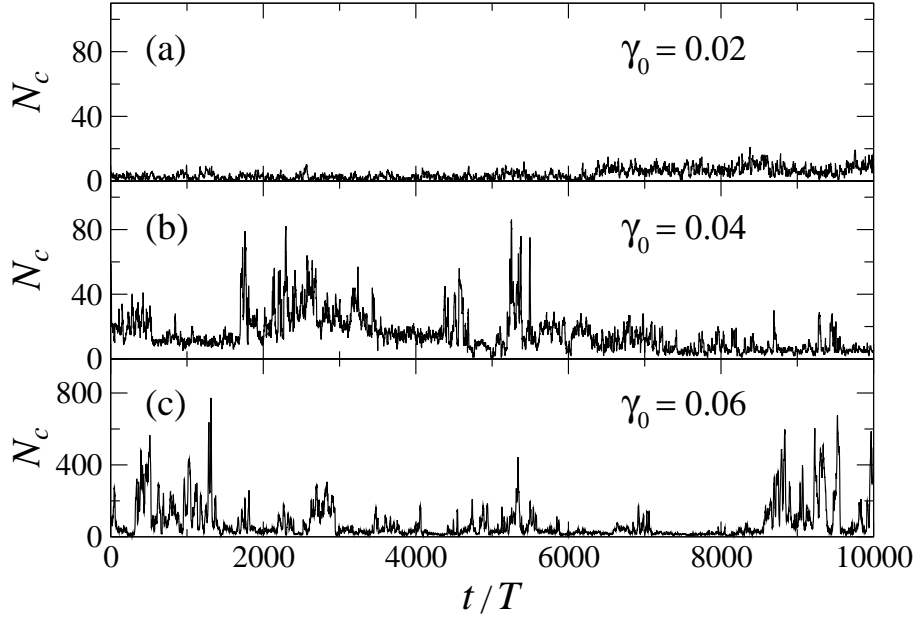}
\caption{(Color online)  The number of  particles undergoing cage
jumps as a function of time for the oscillation frequency
$\omega\tau=0.02$, period $T=2\pi/\omega=314.16\,\tau$, and strain
amplitudes (a) $\gamma_{0}=0.02$, (b) $\gamma_{0}=0.04$, and (c)
$\gamma_{0}=0.06$.   Note that the vertical scale is different in
the panel (c).   } \label{fig:time_clusters}
\end{figure}

Figure\,\ref{fig:time_clusters} shows the total number of particles
undergoing cage jumps as a function of time for three representative
cases at strain amplitudes $\gamma_{0}=0.02,~0.04,~\text{and}~0.06$.
It is clearly observed that the periodic deformation generates a
heterogeneous temporal response characterized by intermittent bursts
of large particle displacements.   It is apparent that the amplitude
of the bursts and frequency of their occurrence increase at larger
strain amplitudes.    During the time intervals between the bursts,
we also detect a finite number of cage jumps that are assisted by
thermal activation.    Following tradition, the frequency spectrum
of the data series in Fig.\,\ref{fig:time_clusters} was determined
by computing their discrete Fourier transform.   Within the reported
time interval, the power spectrum at each strain amplitude exhibits
a power-law decay with the exponent of about two (not shown), which
is indicative of a simple Brownian noise.  This is in contrast with
the inverse frequency spectrum of the so-called flicker noise found
in many complex systems that are characterized by scale-invariant
avalanche-like processes and described by the phenomenon known as
self-organized criticality~\cite{Bak87,BakNature,Vollmayr06}.

\begin{figure}[t]
\includegraphics[width=12.cm,angle=0]{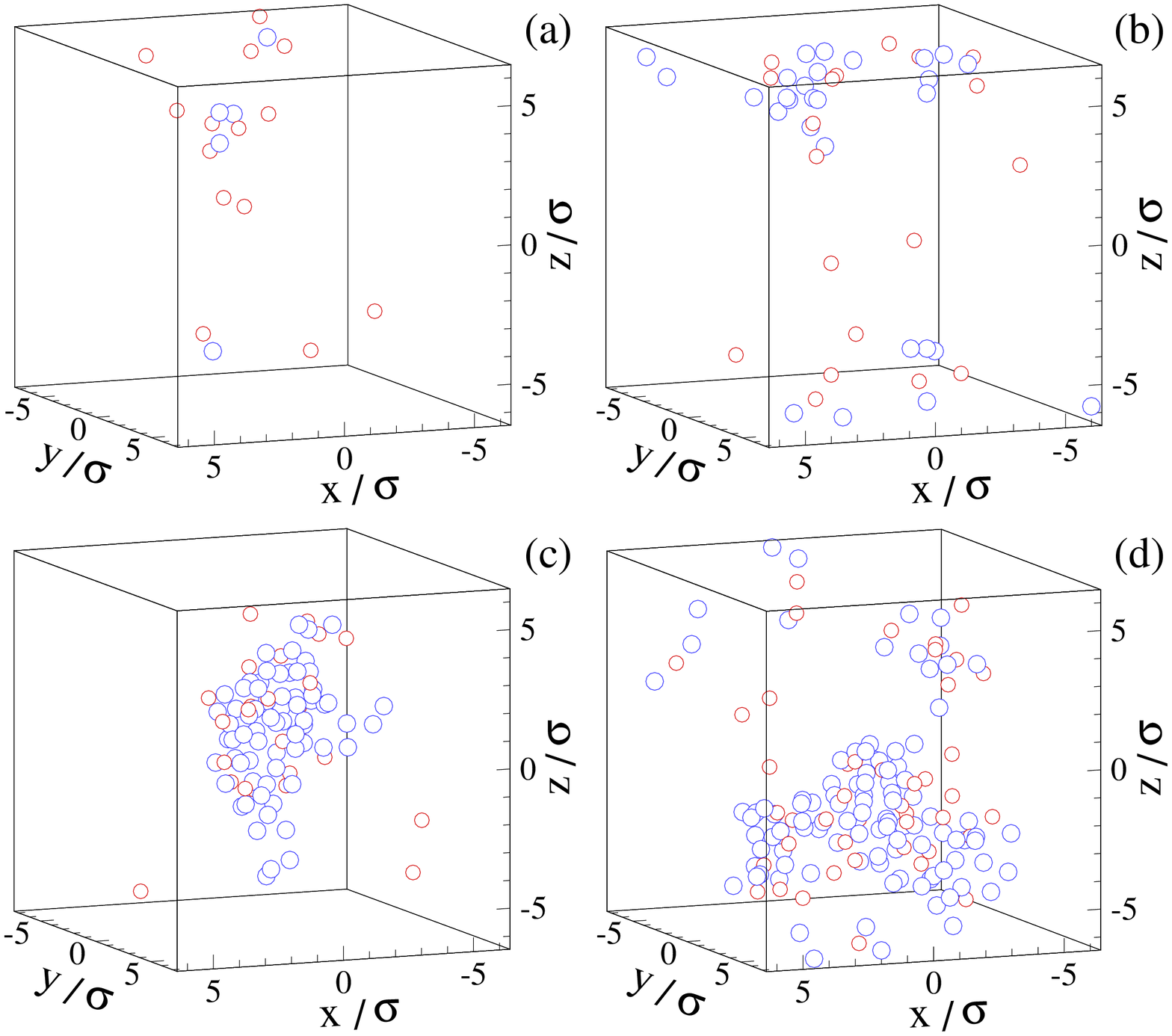}
\caption{(Color online)  Typical clusters of mobile particles $A$
(blue circles) and $B$ (red circles) for the oscillation frequency
$\omega\tau=0.02$ and strain amplitudes (a) $\gamma_{0}=0.02$, (b)
$\gamma_{0}=0.03$, (c) $\gamma_{0}=0.04$, and (d) $\gamma_{0}=0.05$.
} \label{fig:snapshot_clusters}
\end{figure}

A more direct evidence of spatial heterogeneity can be obtained from
visualization of instantaneous positions of mobile particles.
Snapshots of mobile particle positions during intermittent bursts
are presented in Fig.\,\ref{fig:snapshot_clusters} for different
strain amplitudes.   It is clearly seen that the particles
undergoing cage jumps mostly aggregate into clusters whose sizes
increase at larger strain amplitudes.    A number of previous
studies of heterogeneous dynamics in glassy materials have
demonstrated that the sizes of clusters of mobile particles are
power-law distributed~\cite{Vollmayr06,BiroliPRL09,BiroliEPL10}. As
shown in Fig.\,\ref{fig:time_clusters}, the structural relaxation
process involves only a few large-scale cooperative clusters during
the time interval $10^4\,T$, and, as a consequence, we find that the
probability distribution of cluster sizes of more than about $20$
particles is subject to large statistical uncertainty (not shown).

It was recently suggested that dynamic facilitation might be one of
the important mechanisms leading to spatio-temporal heterogeneity in
glassy materials~\cite{Chandler03}.   In the context of kinetically
constrained models, dynamic facilitation can be quantified either by
the mobility transfer function between highly mobile regions and
nearby regions that were previously mobile, or by the facilitation
volume, which measures the spatial extent of mobile regions
initiated by localized excitations~\cite{Keys12}.   In MD
simulations of glass-forming liquids, the mobility transfer function
was computed for mobile particles near their neighbors that were
previously mobile~\cite{Glotzer04,Glotzer05}.    In particular, it
was demonstrated that mobility propagates continuously through the
system, and dynamic facilitation becomes increasingly pronounced
upon supercooling~\cite{Glotzer04,Glotzer05}.

\begin{figure}[t]
\includegraphics[width=12.cm,angle=0]{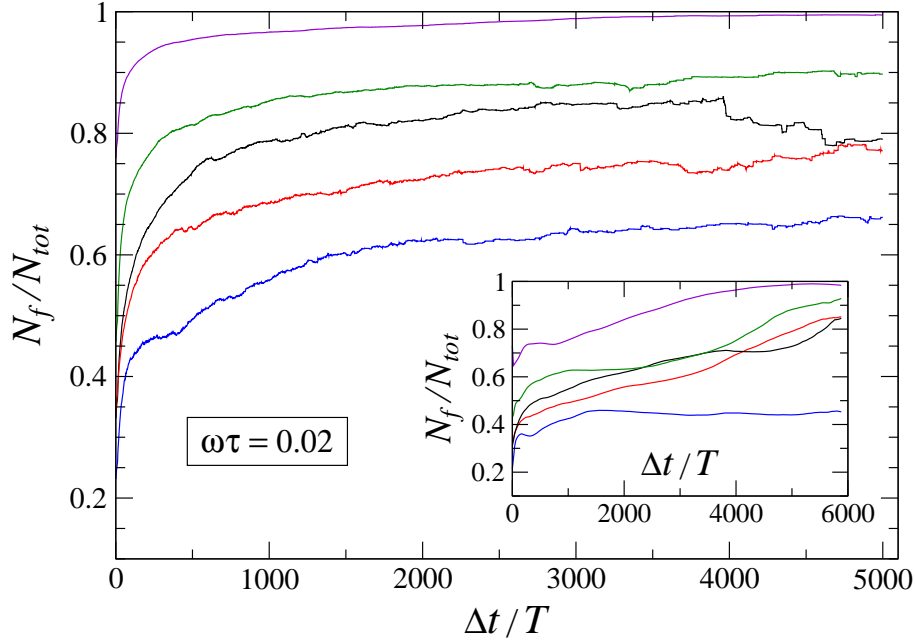}
\caption{(Color online)   The ratio of dynamically facilitated
mobile particles and the total number of mobile particles at a given
time step, provided that they were immobile during the preceding
time interval $\Delta\,t$ (see text for details).    The strain
amplitudes from bottom to top are
$\gamma_{0}=0.02,~0.03,~0.04,~0.05,~0.06$.   The inset shows the
ratio of dynamically facilitated particles and the total number of
particles that become mobile during the time interval $\Delta\,t$,
given that they were immobile during the previous time interval
$\Delta\,t$.   The color code for $\gamma_{0}$ is the same. }
\label{fig:dyn_facil}
\end{figure}

In the present study, two measures of dynamic facilitation were
considered based on the results of the cage detection algorithm
applied to individual particle trajectories.    First, it was
determined whether a particle was mobile at a given time step and it
remained immobile during the preceding time interval $\Delta\,t$.
Next, we checked if the particle had at least one mobile nearest
neighbor during the time interval $\Delta\,t$.
Figure\,\ref{fig:dyn_facil} shows the ratio of dynamically
facilitated mobile particles and the total number of particles that
become mobile after $\Delta\,t/T$ cycles.   It can be seen that the
ratio $N_{f}/N_{tot}$ increases with increasing strain amplitude,
implying that dynamic facilitation plays a more important role at
larger strain amplitudes.   Note also that at the strain amplitude
$\gamma_{0}=0.06$ and $\Delta\,t/T \gtrsim 10^3$, nearly all
particles undergo cage jumps after being in contact with mobile
regions.   In contrast, at the strain amplitude $\gamma_{0}=0.02$,
the ratio $N_{f}/N_{tot}$ appears to saturate at about $0.65$, which
characterizes the relaxation dynamics that involves single particles
undergoing reversible jumps and rearrangement of small clusters of
particles [e.g., see Fig.\,\ref{fig:snapshot_clusters}\,(a)].

Similar to the analysis presented in the previous MD
studies~\cite{Glotzer04,Glotzer05}, we also computed the correlation
between particle mobilities in back-to-back time intervals of equal
duration  $\Delta\,t$.    Namely, we only selected particles that
were mobile at least once during the time interval $\Delta\,t$ but
always immobile during the preceding time interval $\Delta\,t$.
Then, the dynamically facilitated mobile particles were identified
if there was at least one mobile nearest neighbor during the
preceding time interval $\Delta\,t$.   The results are shown in the
inset of Fig.\,\ref{fig:dyn_facil}.   Although the data are somewhat
noisy, the trend is clear; an increasingly larger fraction of mobile
particles are dynamically facilitated at larger strain amplitudes.
Thus, regardless of the definition, the results of numerical
simulations indicate that, as the strain amplitude increases, there
is a higher probability to find a mobile particle that was
previously located near mobile regions.

\section{Conclusions}

In this paper, we performed molecular dynamics simulations to study
heterogeneous relaxation dynamics in an amorphous material under
time-periodic shear strain deformation.      The three-dimensional
amorphous material was modeled as the binary Lennard-Jones mixture
at a temperature well below the glass transition.        During the
oscillatory deformation, the particle positions were stored every
cycle when the net strain is zero.      We found that at small
strain amplitudes, the mean square displacement develops an extended
sub-diffusive plateau followed by the diffusive regime; whereas at
larger amplitudes only the diffusive regime is present at the
reported time scales.

The structural relaxation was described by the decay of the
self-overlap correlation function, which indicated that at small
strain amplitudes the system dynamics is nearly reversible over
about $10^4$ cycles, while at strain amplitudes above a few percent,
almost all particles undergo irreversible displacements and escape
their cages.    With increasing strain amplitude, the dynamic
susceptibility exhibits a pronounced peak at intermediate time and
length scales, and the magnitude of the peak increases at larger
strain amplitudes, indicating progressively larger size of
dynamically correlated regions.    Furthermore, the detailed
analysis of particle hopping dynamics revealed that the periodic
deformation generates a heterogeneous temporal response
characterized by intermittent bursts of large particle
displacements.    Lastly, our numerical simulations have shown that
dynamic facilitation of mobile particles becomes increasingly
important as the strain amplitude increases.

\section*{Acknowledgments}

Financial support from the National Science Foundation
(CBET-1033662) is gratefully acknowledged. Computational work in
support of this research was performed at Michigan State
University's High Performance Computing Facility.



\bibliographystyle{prsty}

\end{document}